\documentclass{aa}

\usepackage{graphicx}
\usepackage{txfonts}
\usepackage[colorlinks=true, linkcolor=blue, citecolor=blue, urlcolor=blue]{hyperref}
\usepackage{longtable}
\usepackage{appendix}
\usepackage{caption}
\usepackage{lastpage}
\usepackage{threeparttable}
\usepackage{comment}
\usepackage{color}
\usepackage{amsmath}
\usepackage[dvipsnames,table]{xcolor}

\newcommand\newtext[1]{{#1 \sl}}

\begin{document}

   \title{Early intrafamily collisions in newly formed asteroid families}

   \author{R. Balossi\inst{1}
          \and 
          P. Tanga\inst{1}
          \and
          A. Dell'Oro\inst{2}
          }

    \institute{Université Côte d'Azur, Observatoire de la Côte d'Azur, CNRS, Laboratoire Lagrange, Bd de l'Observatoire, CS 34229, 06304 Nice Cedex 4, France\\
   \email{roberto.balossi@oca.eu}
   \and
             INAF, Osservatorio Astrofisico di Arcetri, 50125 Firenze, Italy
   }
   \date{}
 
  \abstract
   {The dynamical and physical properties of asteroid family members are widely used to reconstruct the collisional evolution of the main belt and of individual objects. Asteroid families offer insights into the properties of the parent bodies and the fragmentation processes responsible for their formation.} 
   {We investigate a poorly constrained phase of early collisional evolution among members of the same family. Our goal is to determine an intrinsic collision probability associated with intrafamily collisions and to assess their relevance compared to collisions with the background asteroid population.}
   {We performed numerical simulations of the early dynamical evolution of asteroid families, up to the randomization age of the true anomalies, recording mutual impacts between family members and converting them into an intrinsic collision probability. This probability was then used to study intrafamily collisions for generic size distributions.}
  {We identified an intense phase of low-velocity intrafamily collisions occurring in the first few years after family formation. The intrinsic collision probability can reach values up to $10^{-10}\,\mathrm{yr^{-1}\,km^{-2}}$ shortly after breakup and then decreases exponentially, following the same temporal trend predicted by previous statistical models. Variations among the orbital elements of the parent body and the properties of the ejection velocity field can change the collision probability by up to one or two orders of magnitude, without affecting its temporal evolution. Depending on the assumed size distribution, the number of impacts on the largest remnant ranges from fewer than ten to several million.} 
  {Intrafamily collisions represent a physical mechanism whose importance must be assessed on a case-by-case basis. Although they are not expected to produce further fragmentation within asteroid families, they might contribute to early surface and structural evolution in some cases, while being negligible in others.}

   \keywords{small bodies -- asteroid families -- numerical simulations}

    \authorrunning{R. Balossi, P. Tanga, A. Dell'Oro}
    
   \maketitle

\section{Introduction} \label{Introduction}

The dynamical and physical properties of asteroid family members are commonly exploited to reconstruct the collisional history of the currently observable main belt and the evolution of individual objects. In fact, families offer an insight into the properties of the parent bodies at different degrees of shattering and into the mechanical properties of the material under strong stress. In this article, we explore a poorly known phase of early collisional evolution among members of the same family.

Fragments produced from collisions may experience an early reaccumulation phase, in which gravitational interactions among ejecta with low relative velocity lead to the formation of rubble-pile aggregates and sometimes bilobate or contact-binary shapes (\citealt{michel-2002, michel-2004, michel-2015, sugiura-2018}). This phase, lasting typical timescales of several hours, generates the original properties of the family members.

After formation, families evolve through dynamical processes, such as the Yarkovsky-driven drift and interactions with resonances, as well as high-velocity collisions with other main belt asteroids. These impacts erode families with a size-dependent intensity, reducing the steepness of the cumulative size distributions. They also generate craters, alter the original surfaces, and create sub-families (\citealt{o'brien&greenberg-2005}, \citealt{bottke-2005}). 

In this work, we assume that the initial, family-forming reaccumulation is complete and we focus on the  phase immediately following, in which the freshly born family members occupy a small volume in the orbital element space (in fact, the smallest possible in the family history). Soon after the family formation, its members disperse with small relative velocities and they share not only the closest values of semi-major axis, eccentricity, and inclination ($a$, $e$, $i$), but also similar node longitudes and argument of perihelia ($\Omega$, $\omega$). The true anomaly, $f$, is also restricted to a limited range of values. 

In such conditions, intrafamily collisions can occur. We can postulate that they should be more frequent at the common nodes of the orbits, where the family members come back nearly at the same time, thanks to a small dispersion in $a$ (and thus, in the orbital periods). One of the nodes will approximately be at the location of the breakup event and the other at the opposite orbital longitude. The gradual increase of dispersion (in the anomalies first) rapidly reduces the frequency of mutual impacts. 

This early collisional phase has received relatively limited attention in the literature. The first preliminary assessment was carried out by \cite{davis-1996}, who predicted that such collisions should not be frequent enough to significantly contribute to the collisional aging of the family size distribution. A more detailed investigation was later performed by \cite{delloro-2002}, who computed the intrinsic collision probability for collisions among family members using a dedicated statistical approach.

Classical formalisms (\citealt{wetherill-1967, greenberg-1982}) cannot be applied in this context because they aim to evaluate the statistical parameters of impact averaging over all possible relative orbital orientation assumed equally probable, an assumption that breaks down during the first 10$^4$-10$^5$ years after family formation, when the orbital angles are not yet randomized. To overcome this limitation, \cite{delloro-2002} developed their own statistical approach, based on the analytical formalism described in \cite{delloro&paolicchi-1998}, valid even when the longitudes of nodes and perihelia are not fully randomized. They found a strong enhancement in the mutual collision rate after family formation, compared to the constant background rate.

\cite{delloro-2002} concluded that this increased collision rate would not change the size distribution of the family, while also stressing the fact that numerical simulations would be required to study the very early phases (of a duration of some $\sim$100s years), when the anomalies are not fully randomized either and the collisional rates could be even stronger.

In this work, we fill that gap by performing numerical simulations of intrafamily collisions to model the collisional evolution at those early phases, thereby achieving a complete scenario. In the process, we also compare our results to the theoretical model of \citet{delloro-2002} to check consistency. Possible applications of our results may concern different aspects of family evolution:
\begin{itemize}
    
    \item Intrafamily impacts in this regime should occur at very low relative velocities. Previous works have largely focused on high-velocity impacts (e.g., \citealt{petit&farinella-1993}), whereas low-velocity collisions have received comparatively little attention. Nonetheless, laboratory experiments (\citealt{uehara-2003}, \citealt{takita&sumita-2013}, \citealt{hayashi-2017}) and numerical simulations (\citealt{celik-2022}, \citealt{langner-2025}) demonstrate that even low-velocity collisions can produce shallow craters, whose properties depend on the impactor and the target compositions. These results suggest that intrafamily collisions may play a non-negligible role in shaping family members, particularly regarding their cratering record or the early development of a surface regolith layer.
    
    \item Asteroid families are commonly assumed to be compositionally homogeneous (\citealt{parker-2008, balossi-2024, balossi-2025}). However, thermal evolution models and meteoritic evidence indicate that many early planetesimals differentiated into core, mantle, and crust layers (e.g., \citealt{weiss&elkins-2013}). Hence, their disruption should have produced compositionally heterogeneous families, yet fragments sampling differentiated interiors are far rarer than expected (e.g., \citealt{burbine-1996}) and recent discoveries such as the olivine-rich family of \cite{galinier-2024} remain exceptions. A possibility (that remains to be verified) is that in certain conditions, post-formation collisions could be particularly efficient in homogenizing family surfaces by hiding compositional diversity.

    \item Intrafamily collisions at low relative velocity by fragments coming from a heterogeneous parent body could support the existence of some spectral variations, especially those requiring gentle merging of similar-size fragments, such as the hemispheric variations found by \citet{hasegawa+2024} in some asteroids of size R$\sim$5--10~km. 
    
\end{itemize}

This short list is not exhaustive and it does not imply that intrafamily collisions are relevant for all asteroid families. In this article, we limit our investigation to the statistical properties of this mechanism as a function of several parameters describing the dispersion of the initial family. We tentatively explore its application to a simple scenario of compositional mixing. 

The article is organized as follows. Section \ref{Methods} describes the numerical simulations and the computation of the intrinsic collision probability. In Sect. \ref{Numerical simulations} we illustrate the fundamental properties of early intrafamily collisions, in comparison to \cite{delloro-2002}. We  examine how the collision probability depends on the ejection-velocity field and on the orbital elements of the parent body at breakup. Section \ref{generalization to real asteroid families} applies the results to arbitrary size distributions of family members. In Sect. \ref{Importance}, we discuss how intrafamily collisions may affect family member surfaces and we illustrate an application of the model. Finally, Sect. \ref{Conclusion} presents our conclusions and perspective for future work.

\section{Methods} \label{Methods}

\newtext{Our first objective is to derive the intrinsic probability, $P_T$, of intra-family collisions. Overall, $P_T$ is a convenient tool since, by definition, it is independent from the size of particles involved in the impacts. It can thus be easily computed by numerical simulations that are limited in particle number. Once available, it can be exploited to compute the absolute collision rates in the case of a full distribution of particle sizes (Sect.~\ref{generalization to real asteroid families}).} Our numerical simulations exploit the N-body code \textit{Rebound} \citep{rein&liu-2012} and its capacity to detect collisions in a system of particles of finite size. 

\subsection{Initial distribution of parent body fragments} \label{fragmentation of the parent body}

We considered fragments produced from a family-forming event, as represented by spheres with radii between a minimum, $r_{min}$, and a maximum value, $r_{max}$. Here, $r_{min}$ is on the order of a few meters, while $r_{max}$ is a fraction of the parent body’s radius. Initially, they are  randomly placed within the parent body's spherical volume. Fragments might overlap, but this does not affect the outcome of the simulations, as collisions occurring during the initial evolution are not recorded. 
Indeed, our investigation focuses on what happens after the ballistic phase,  which is when the initial mutual collisions and gravitational re-accumulation occur (\citealt{michel-2002, michel-2004, michel-2015, jutzi-2019}). In this respect, the initial velocities of the fragments at the beginning of our numerical integration must be understood as the velocities "at infinity" that characterize the fragments  after they leave their mutual spheres of influence, moving along fully independent orbits from that point forward.

To balance the computational cost with statistical reliability, the number of fragments is kept relatively low (i.e., on the order of a few tens of thousands), which naturally reduces the number of collisions during the system’s dynamical evolution. As explained above, without any consequences for the intrinsic collision probability, fragment diameters can be artificially scaled by a size factor of $F_s$, increasing the collision cross-sections and therefore the number of recorded impacts. Since our first goal is to derive an intrinsic collision probability (Sect.~\ref{collision probabilities}), the exact particle sizes are not relevant, as they only act as normalization factors in the probability computation. For this reason, several simulations are run with equal-sized particles.

We also track the dependence of the collisions on the original location of the fragments by dividing the parent body into shells and assigning to each particle a label representing the depth of the layer. These may also be used as a proxy of a radial compositional gradient inside the parent body, or to match layers of a fully differentiated object with core, mantle, and crust. \newtext{In the following numerical simulations (Sect. \ref{Numerical simulations}), we subdivide the parent body radially into three concentric shells, $d < 50$ km, $50 < d < 85$ km, and $85 < d < R_{PB}$, where $d$ is the distance from the center of the parent body of radius $R_{PB}$. These layers can be interpreted as an indicative representation of a core, mantle, and crust for a differentiated parent body and will be used to track their collisional mixing.}

\subsection{Ejection velocity field} \label{Ejection Velocity Field}

\begin{figure}
   \centering
   \includegraphics[width=\hsize]{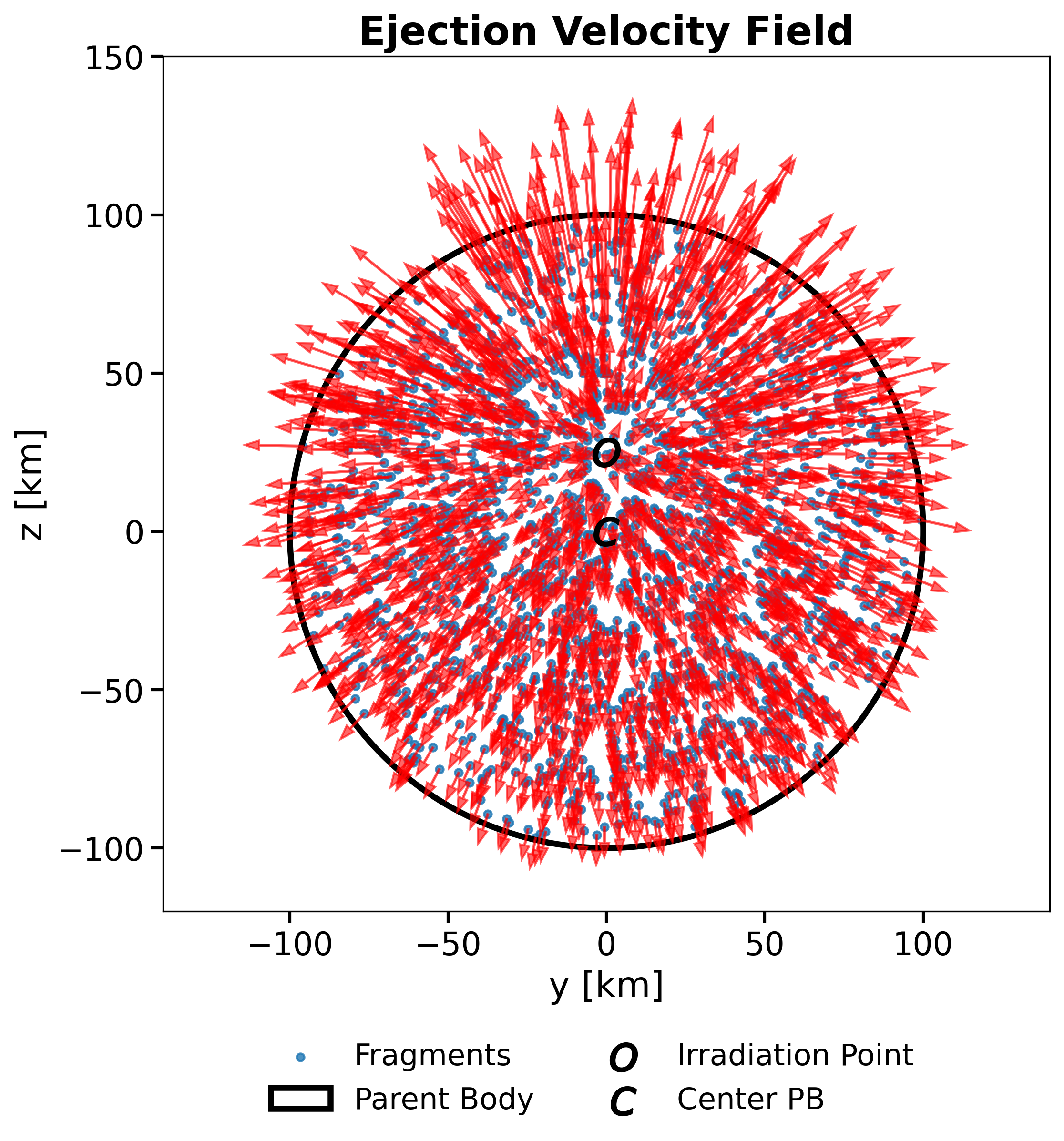}
      \caption{ Example of an ejection velocity field seen in the (z, y) plane. The parent body profile is shown in black, the positions of the fragments in blue, and the velocities in red. The letters O and C mark the irradiation point and the center of the parent body, respectively.}
         \label{plot: ejection velocity field}
\end{figure}

The choice of the initial ejection velocity field (EVF), which determines the dispersion of the orbits of the family members, is relevant to our problem. In this respect, several options are possible. We notice that most (if not all) recent simulations of the dynamical evolution of asteroid families adopt a perfectly isotropic EVF, representing the simplest possible choice. Full SPH simulations of catastrophic disruptions could lead to a complex EVF, potentially more realistic, but the added complexity and variety of configurations do not match our goal of a first general study with a limited number of parameters. \newtext{Also, for the same reason, we want to avoid EVF that are strongly anisotropic, since the preferential direction in the ejection may affect the statistics of collisions. Such situations, mostly due to cratering events, tend to be too specific and less general.}

We thus decided to adopt the representation of the EVF of intermediate complexity, provided by the Semi-Empirical Model (SEM) by \cite{paolicchi-1996}, \newtext{to reproduce fragmentation families. This solution introduces a tunable anisotropy, and it has been validated against the results from laboratory hyper-velocity experiments \citep{giblin-1998}}. As the velocity dispersion is the main factor controlling the impact rate during the subsequent evolution, we can consider that details of the EVF, not present in the SEM, are second-order effects. For instance, at this stage, we neglect more complex correlations between the particle position and velocity, which have been shown to introduce differences with respect to the SEM in Smooth Particle Hydrodynamics (SPH) simulations of catastrophic impacts \citep{delloro-2015}.

The EVF produced by the SEM is radial, with an irradiation point located within the target. At this stage, we neglect the parent body rotation. Therefore, the velocity field in a generic point $P$ within the target will be given by

\begin{equation}
    \vec{V}(P) = V_{B} (R_{PB}/l)^{\gamma}  (d/l)^{\delta} \mathbf{\hat{QP}},
\end{equation}

where $R_{PB}$ is the radius of the parent body, $d$ is the distance of $P$ from the irradiation point $Q$, $\mathbf{\hat{QP}}$ is the unit vector along the direction $QP$ and $l$ is the length of the segment connecting the irradiation point to the surface passing in $P$ (for a visual representation, see Fig. 1 in \citealt{paolicchi-1996}). $V_B$ is the value of the module of the velocity field for a point on the surface and at a
distance $R_{PB}$ from the irradiation point. The exponents $\gamma$ and $\delta$ control the rate of change of velocity with respect to the position in the target, relative to $Q$ and $P$. The point $Q$ is assumed to be at a distance $S$ from the center of the parent body, along the z-axis. The value of $S$ is parametrized by $\phi = 1 - S/R_{PB}$. Therefore, $\phi=1$ describes a breakup whose EVF is isotropic and with the irradiation point located at the center of the parent body, while $\phi = 0$ implies that the irradiation point is located at the surface.

$V_B$ is constrained following \cite{carruba-2003}. Given the density and the radius of the parent body, $\rho_{PB}$ and $R_{PB}$, the specific energy of the parent body's fragmentation is computed from $Q_D = 0.4 \rho_{PB} (R_{PB})^{1.36}$, where $Q_D$ corresponds to a 0.5 mass ratio of the largest remnant to the parent body, which conventionally defines the transition between the cratering and catastrophic disruption regimes \citep{benz&asphaug-1999}. Different parametrizations can then be adopted to reproduce a variety of families. Assuming that only a fraction $f_{KE}$ of $Q_D$ is converted into kinetic energy, then $V_B = \sqrt{Q_D f_{KE}}$, where $f_{KE}$ is usually taken between 0.01 and 0.1. 

An example of an ejection velocity field for 2,000 particles is reported in Fig. \ref{plot: ejection velocity field}. \newtext{This EVF was generated assuming a parent-body density and radius of $\rho_{PB} = 2$ g cm$^{-3}$ and $R_{PB} = 100$ km. The exponents were set to $\gamma = 3.8$ and $\delta = 1.2$, typical values from \cite{paolicchi-1996}. The position of the irradiation point is parametrized by $\phi = 0.75$, representing a typical case for a fragmentation family as measured in hypervelocity impact experiments \citep{giblin-1998}. Finally, we adopt $f_{KE} = 0.01$ to keep the initial family distribution as compact as possible, which yields $V_B = 51.47$ m s$^{-1}$. These parameters are used in the following numerical simulations unless stated otherwise.}

\newtext{The velocity distribution produced by the SEM is broadly similar to a Maxwellian distribution, with a tail extending toward higher velocities. The exact shape depends on the parameters adopted in the model. For the EVF shown in Fig. \ref{plot: ejection velocity field}, the minimum velocity is on the order of 1 m/s, the maximum velocity is about 150 m/s, and the distribution peaks around 30 m/s.}

The EVF described by the SEM is divergent. For fragments in the same radial direction, the velocity increases with the distance from the irradiation point. Fragments closer to the surface will move faster than those near the center, preventing any collisions from occurring immediately after breakup. 

\subsection{Integration} \label{Integration of the System}

Once the fragments are generated within the parent body and assigned an ejection velocity field, the system is integrated over time, and collisions are recorded. 

The system is integrated using the N-body integrator \textit{Rebound} \citep{rein&liu-2012} with the symplectic Wisdom-Holman integrator WHFAST \citep{rein&tamayo-2015} and a timestep of 0.01 yr. The dynamical model includes the gravitational perturbations of the Sun and the eight major planets, whose initial positions and velocities were obtained from JPL Horizons at the start of each integration. Family members evolve under these gravitational forces as well as non-gravitational perturbations, in particular the Yarkovsky effect and solar radiation pressure, which are included via the \textit{Reboundx} package (\citealt{tamayo-2020}, \citealt{ferich-2022}). We verified that, over the short integration timescales considered (a few hundred years), one-meter fragments spread due to the Yarkovsky effect by an amount smaller than the spatial dispersion caused by planetary gravitational interactions. Smaller particles disperse more rapidly, typically over a few orbital periods of the parent body, so their contribution to collisions is limited. For these reasons, we set the minimum fragment size in our simulations to one meter. This represents a reasonable compromise between computational cost and the total number of collisions detected, and excluding smaller particles should not significantly affect the resulting intrinsic collision probability.

\newtext{Gravitation interactions among family members are not considered at this stage. In doing so, we assume that gravitational scattering by the largest family members is affecting only a negligible minority of small fragments. We are aware that this simplification may subtly alter some results, but since our goal here is to start investigating the main features of the early intrafamily collisional phase, we leave the analysis of a more complex scenario to future studies.}

Collisions are recorded during the integration using the detection module $line$, already implemented in \textit{Rebound}. This algorithm is a brute force collision search, which scales as $O(N^2)$ and checks for overlaps between particles by assuming linear trajectories over each timestep and analytically interpolating their relative motion within the step\footnote{\url{https://rebound.readthedocs.io/en/latest/collisions/}}. When a collision is detected, the time of collision and the colliding particles are saved for later analysis. 

\subsection{Intrinsic collision probability} \label{collision probabilities}

After collisions are recorded, their count can be exploited to compute an intrinsic collision probability following \cite{dahlgren-1998}:

\begin{equation}
    P_T(t) = \frac{N_{coll} (<t)}{n_T \ t \ (R_p^2 + R_t^2) \ F_s^2},
    \label{eq:probability}
\end{equation}
where $N_{coll}(< t)$ is the number of collisions before the time $t$, $R_p$ is the radius of the projectile, $R_t$ is the radius of the target, and $F_s$ is the scale factor used to increase particle sizes, as described in Sect. \ref{Integration of the System}. $n_T$ represents the number of possible fragment pairs. For indistinguishable particles, $n_T = 0.5\ n\ (n - 1)$, where $n$ is the number of particles. It is instead equal to $n_T = n_i n_j$ for distinguishable particles, for instance, when collisions between different shells of the parent body are considered.

\section{Numerical simulations} \label{Numerical simulations}

\begin{figure}
   \centering
   \includegraphics[width=\hsize]{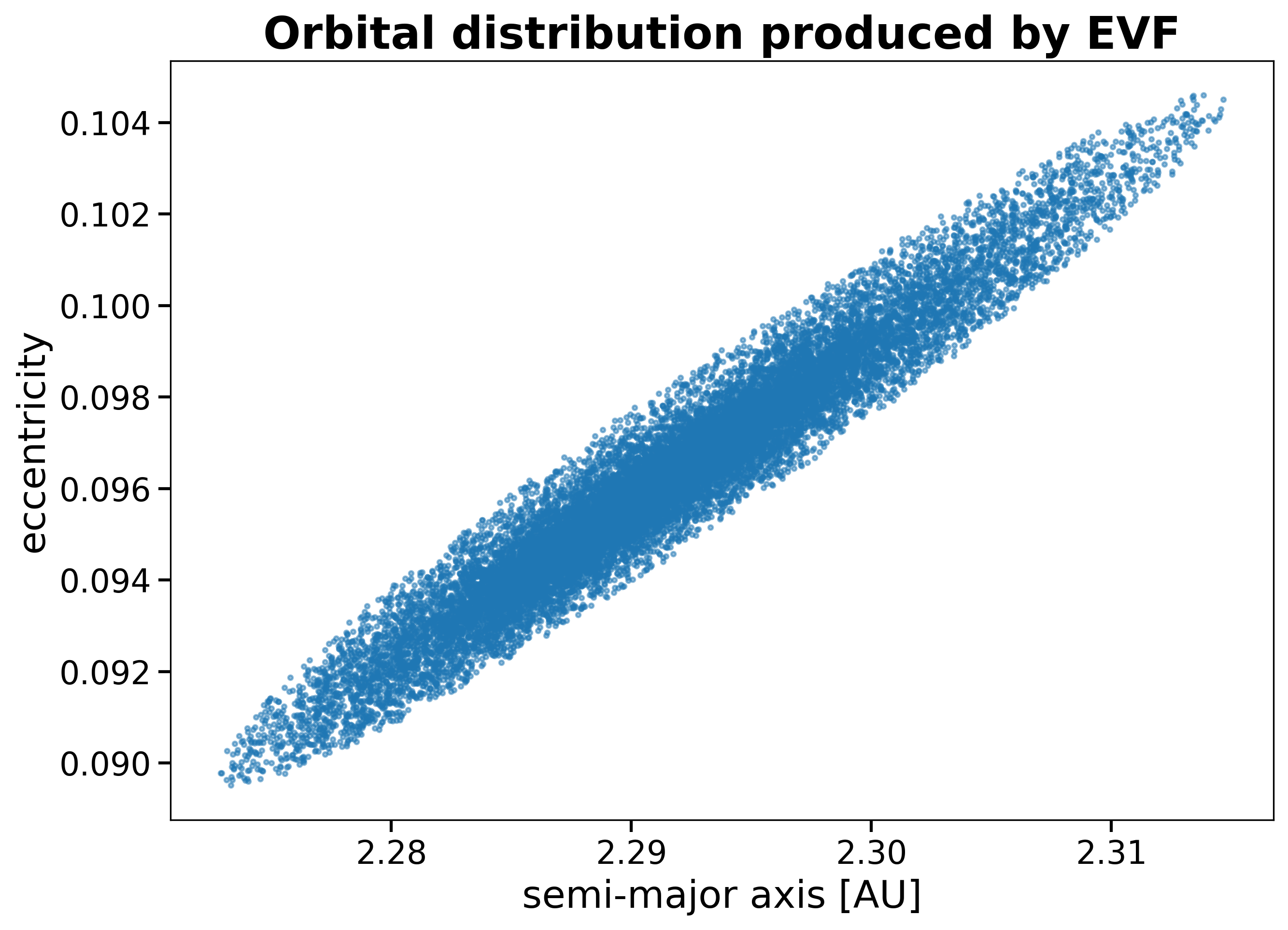}
      \caption{ \newtext{Distribution of the orbital elements in the ($a$, $e$) plane produced by the ejection velocity field.}}
    \label{plot: orbital elements evf}
\end{figure}

\begin{figure*}[ht]
   \centering
   \includegraphics[width=\textwidth]{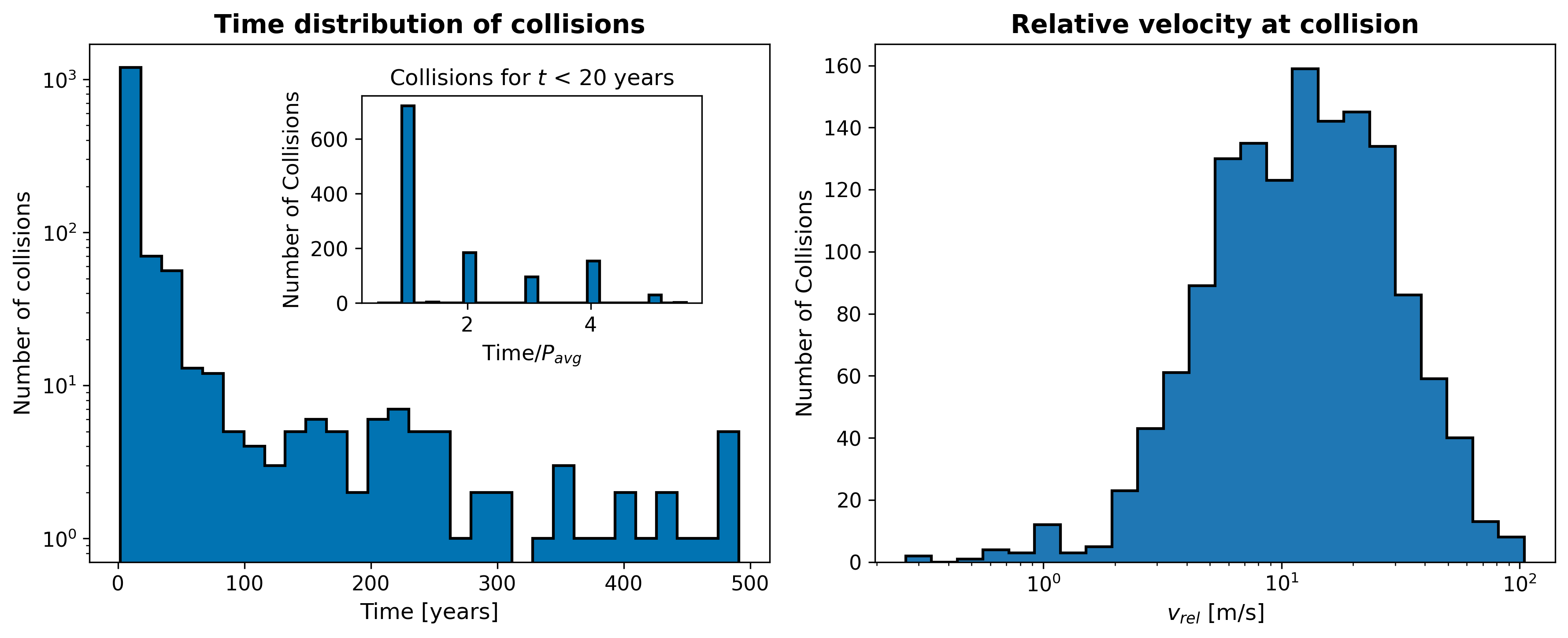}
      \caption{ Details of the collisions recorded during the integration of a system of 25,000 particles, each with a radius of 50 km. Left panel: Temporal distribution of collisions. \newtext{The inset provides a close-up of the first twenty years of the integration. In this panel, collision times are normalized to the average orbital period of the fragments, $P_{avg}$, showing that impacts preferentially occur at multiples of this period.}Right panel: Distribution of impact velocities between the fragments. }
    \label{plot: collisions_integration}
\end{figure*}

In this section we present an example of a numerical simulation and investigate how the intrinsic collision probability depends on the model’s free parameters.

\subsection{Example of a numerical simulation} \label{example of a numerical simulation}

\newtext{We performed several numerical simulations using different parameter sets. We discuss the dependence of the results on selected parameters in Sect. \ref{collision probability dependencies}}. As a representative example, we present the result of the integration of a system of 25,000 particles, each with a radius $r$ = 5 km, artificially scaled by a factor $F_s = 10$, over a period of 500 years. The orbital elements of the parent body at the moment of breakup were assumed to be: semi-major axis $a_{PB} = 2.3$ AU, eccentricity $e_{PB} = 0.1$, inclination $\sin(i_{PB}) = 0.15$, longitude of the ascending node $\Omega_{PB} = 70^\circ$, argument of the perihelion $\omega_{PB} = 30^\circ$, and true anomaly $f_{PB} = 30^\circ$. The EVF is modeled with the parameters reported in Sect. \ref{Ejection Velocity Field}. \newtext{The distribution of the orbital elements produced by the initial ejection velocity field described in Sect. \ref{Ejection Velocity Field} is shown in Fig. \ref{plot: orbital elements evf}. Fragments ejected at lower velocities form a compact, dense cluster, while those ejected at higher velocities create a more dispersed halo.}

The temporal distribution of the collisions recorded during this integration is shown in the left panel of Fig. \ref{plot: collisions_integration}. As expected, collisions are most frequent immediately after family formation and decrease as the fragments disperse in space. \newtext{The inset shows that, at early times, collisions preferentially occur at multiples of the average orbital period of the fragments, $P_{avg}$, which is similar to the parent body's orbital period $P_{PB}$. This behavior arises from the initial geometric configuration of the fragment orbits. Immediately after breakup, the fragments share very similar orbital elements and have clustered true anomalies, so that they periodically return to a similar configuration near the breakup location, enhancing the collision rate. Over the short timescales considered here, the orbital elements evolve only slightly, while the true anomalies rapidly disperse along the orbit. This diffusion drives the decay of the collision rate and causes the periodicity to disappear once the true anomalies become fully randomized.}

\begin{figure}
   \centering
   \includegraphics[width=\hsize]{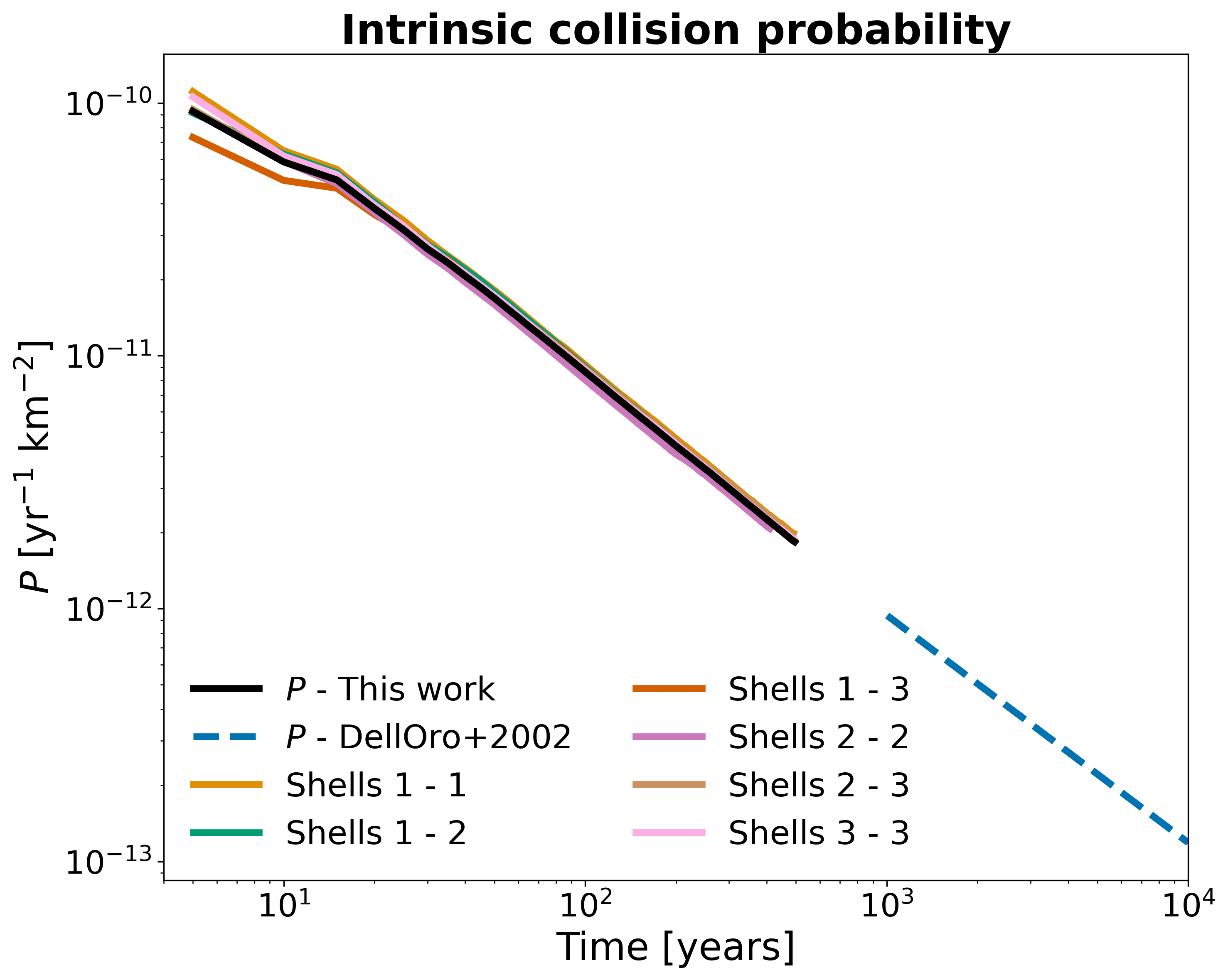}
      \caption{ Intrinsic collision probability derived from the system described in Sect. \ref{Integration of the System}.  The collision probability for the whole system is shown in black. The colored curves represent collision probabilities for fragments originally located in different shells of the parent body. The dashed blue line shows the collision probability reported in \cite{delloro-2002}.}
    \label{plot: collision probability}
\end{figure}

The distribution of relative impact velocities for each colliding particle pair is shown in the right panel of Fig. \ref{plot: collisions_integration}. All collisions occur at very low velocities, ranging from around 1 m/s to a maximum of 100 m/s, with a peak around 15 m/s.

Consequently, intrafamily collisions are not expected to cause significant fragmentation or alter the family’s size distribution, consistent with the findings of \cite{delloro-2002}. Nonetheless, these low-velocity impacts may influence the cratering record on the surfaces of family members. The potential physical effects of these collisions are further discussed in Sect. \ref{Importance}.

\begin{figure*}[ht]
   \centering
   \includegraphics[width=\textwidth]{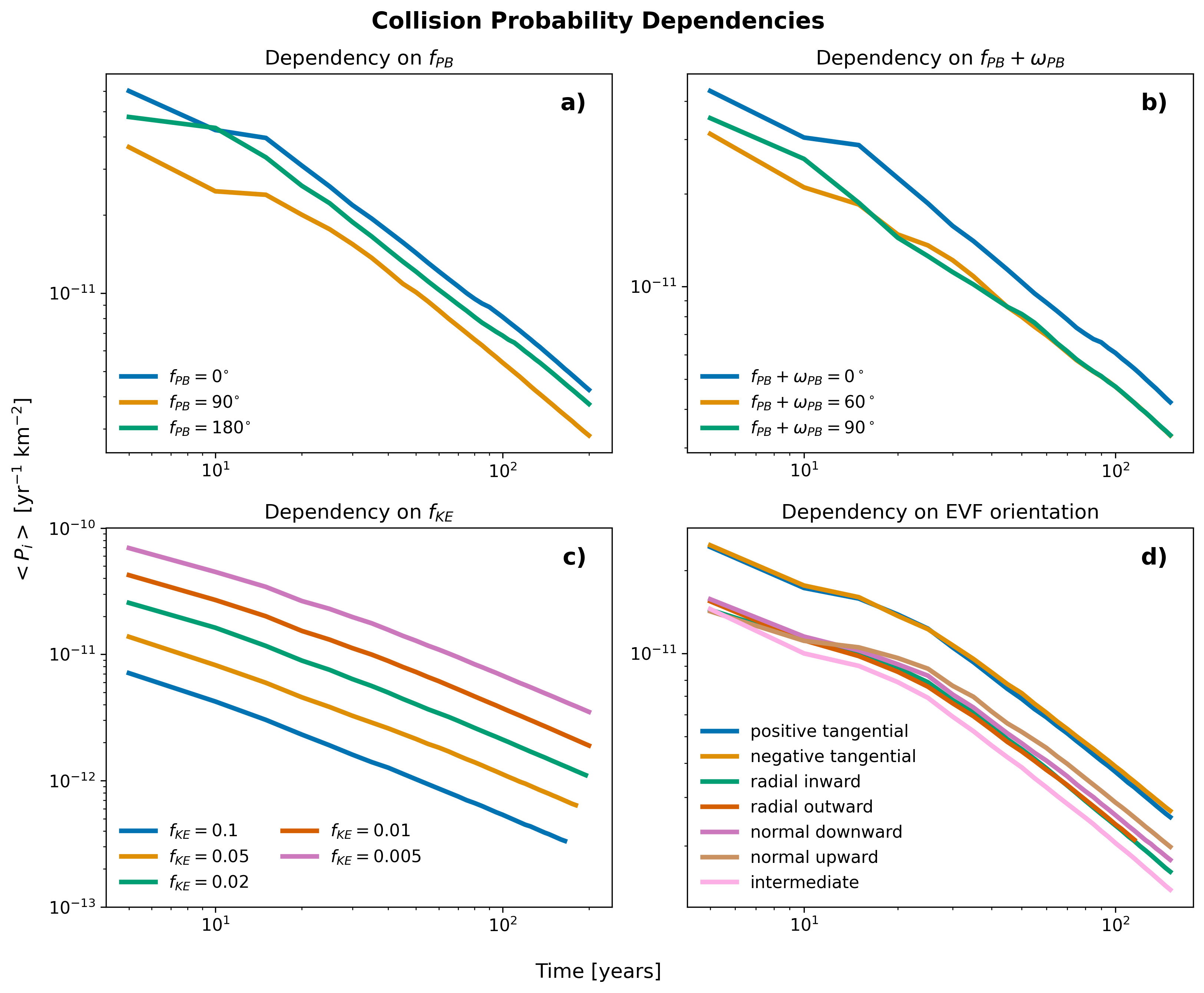}
      \caption{Collision probability as a function of the parent body’s orbital elements at breakup and the ejection velocity field properties. Panel (a): Dependence on $f_{PB}$. Panel  (b): Dependence on $f_{PB} + \omega_{PB}$. Panel (c): Dependence on $V_B$, here parametrized by $f_{KE}$. Panel (d): Dependence on the orientation of the ejection velocity field.  }
    \label{plot: P_dependencies_EVF_PB}
\end{figure*}

\cite{delloro-2002} report that intrinsic collision probability follows a power-law dependence upon the time, which, for $t < t_R$, can be described as
\begin{equation}
    P(t) = P_0 \ (t/t_R)^{-z},
    \label{eq:probability-delloro}
\end{equation}
where $t_R$ is the randomization age of the longitudes of the ascending nodes, $P_0$ is the classical mean intrinsic probability, and $z$ is an exponent that depends on the orbital elements of the parent body. \cite{delloro-2002} find that $z$ is mainly constrained between 0.75 and 0.8, and reaches values up to 0.9 only for particular geometric configurations of the parent body's orbit.

Figure \ref{plot: collision probability} reports the intrinsic collision probability $P$ derived from the integration. $P$ follows an exponential law described by $P(t) = 3.5 \times 10^{-10} \ t^{-0.79}$, showing a trend slope in excellent agreement with \citet{delloro-2002}. Figure \ref{plot: collision probability} also reports an intrinsic collision probability derived from the statistical model and described by Eq. \ref{eq:probability-delloro}. This probability ranges between the randomization age of the true anomalies, assumed to be 1000 years, and the randomization age of the longitude of the ascending nodes, $t_R$, which is computed as $t_R = 180^{\circ}/\sigma_{\dot{\Omega}}$, where $\sigma_{\dot{\Omega}}$ is the standard deviation of the average rate of circulation of the longitude of the node (which we assumed $\sigma_{\dot{\Omega}}$ = 1.5 arcsec/year, taken as a representative value from \citealt{delloro-2002}).
The intrinsic collision probability obtained from the numerical simulations follows the same trend and slope as that obtained by the statistical model, converging to it at later times. This demonstrates that the results obtained by \cite{delloro-2002} can be reliably extrapolated even before the randomization age of the true anomalies. During this phase, the increase in intrinsic probability continues back in time, reaching a factor of two additional orders of magnitude. 

An additional insight in the significance of these results can be obtained by considering a radial subdivision of the parent body into the three concentric shells \newtext{described in Sect. \ref{fragmentation of the parent body}}. We computed the intrinsic collision probabilities for the collisions among the fragments of different shells, which are shown in Fig. \ref{plot: collision probability}. These probabilities closely match the total collision probability. This indicates effective mixing within the parent body, such that particles originating from different shells are just as likely to collide as those from the same shell. The physical implications of this result will be discussed in Sect. \ref{Importance}.

\subsection{Collision probability dependences} \label{collision probability dependencies}

\cite{delloro-2002} noticed that the rate of collisions is enhanced in specific geometric configurations, namely when the true anomaly of the parent body $f_{PB} = 0^\circ$ and $f_{PB} = 180^\circ$, as the fragments share the same perihelion or aphelion, respectively. They also report an enhancement when $f_{PB} + \omega_{PB} = 90^\circ$, as the fragments have the same inclination. Finally, they verify that, as expected, large values of $V_B$ determine a larger spread in the orbital elements space, thus decreasing the collision rate between particles. At the same time, the mean impact velocity at any epoch increases with increasing $V_B$.

We numerically verified the collision probability dependences described by \cite{delloro-2002} and also investigated the effect of the ejection velocity field and of the initial location of the fragments within the parent body. For each case, we integrated a system of 25,000 particles with a radius of 50 km, as described in Sect. \ref{Integration of the System}, testing the effect of changing a single parameter. We also verified that the integration results do not depend on the particle size, as varying the particle radii produces the same collision probability. The results for the orbital elements of the parent body and the EVF are reported in Fig. \ref{plot: P_dependencies_EVF_PB}, while no dependence on the initial configuration of the fragments is found. \newtext{In all the following simulations, we assume that the semi-major axis of the parent body at breakup is $a_{PB} = 2.3$ AU. This value was chosen because it lies sufficiently far from any major resonance. We nevertheless performed additional numerical simulations varying the semi-major axis of the parent body. In all cases, we observed the same temporal evolution of the intrinsic collision probability (scaled to the orbital period). This behavior is expected, since the dynamical environment and the background collision probability become relevant only on long timescales, much longer than those over which intrafamily collisions are effective, which are primarily determined by the parent body's breakup conditions.}

To study the dependence on the true anomaly of the parent body $f_{PB}$ at the time of breakup, we integrated three systems with $f_{PB} = 0, 90, 180 ^\circ$, $a_{PB}$ = 2.3 AU, $e_{PB}$ = 0.1, $\sin(i_{PB})$ = 0.15, $\omega_{PB} = 50 ^\circ$, and $\Omega_{PB} = 0 ^\circ$. To explore the dependence on $f_{PB} + \omega_{PB}$, we again integrated three systems, where $a_{PB}$, $e_{PB}$, and $i_{PB}$ were taken as before, $\Omega_{PB} = 90 ^\circ$, $f_{PB} = 0 ^\circ$, and $\omega_{PB} = 0, 60, 90 ^\circ$. For both cases, the EVF was modeled with the parameters reported in Sect. \ref{Ejection Velocity Field}. Panel (a) shows that for $f_{PB} = 0^\circ$ and $f_{PB} = 180^\circ$ the collision probability is larger than other values of $f_{PB}$, thus validating the results of the statistical approach. Panel (b) instead reports that the collision probability for $f_{PB} + \omega_{PB} = 0^\circ$ lies above the other two. This is compatible with \cite{delloro-2002}, although they report that their results at short time scales are not reliable due to mathematical instabilities.

\begin{figure}
   \centering
   \includegraphics[width=\hsize]{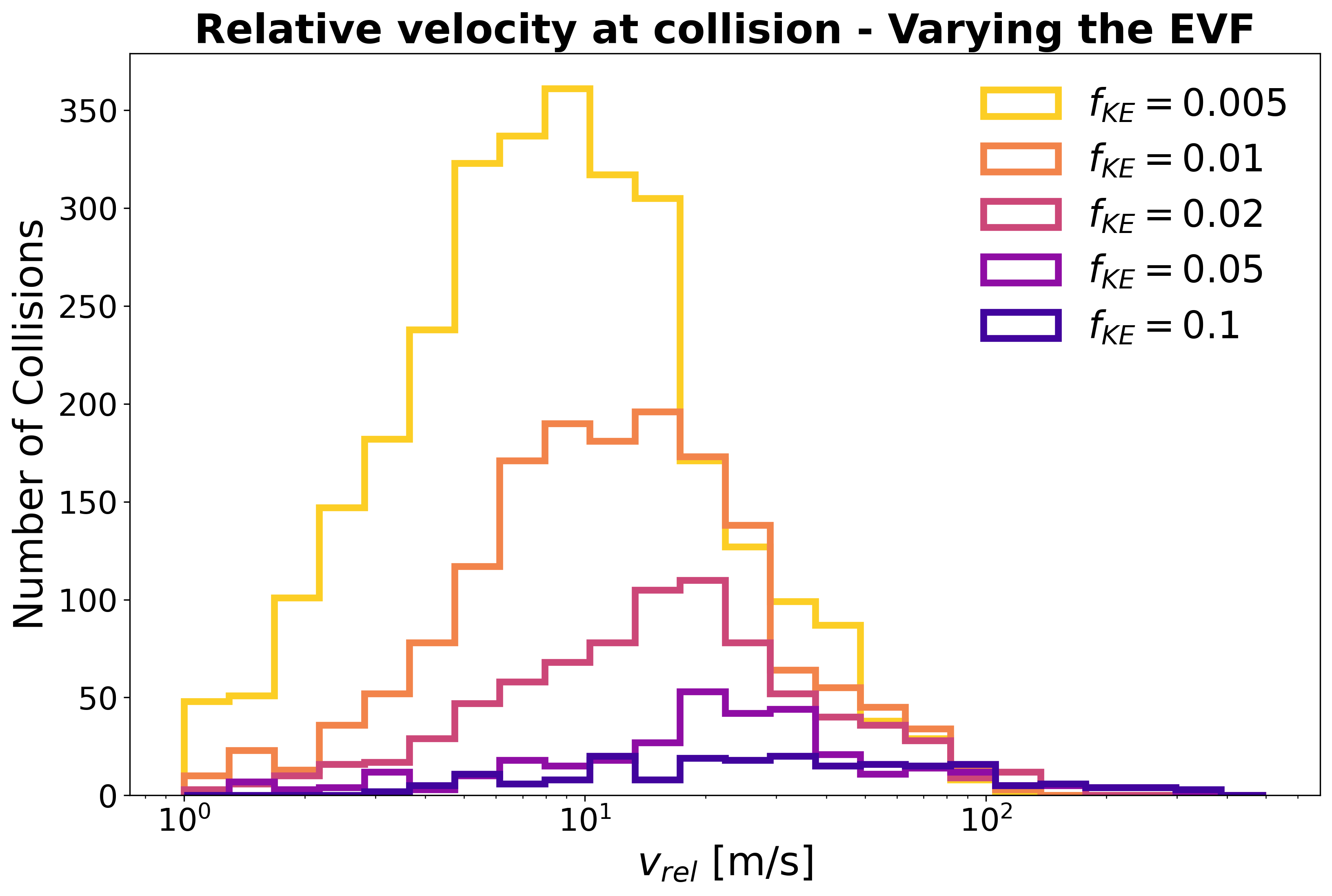}
      \caption{ Distributions of the impact velocities for the same system integrated varying $V_B$, here parametrized by $f_{KE}$.}
    \label{plot: impact velocity EVF dependency}
\end{figure}

We also investigated the effect of $V_B$, which we parametrized by varying $f_{KE}$ (panel c). We took $f_{KE}$ in the range 0.005 -- 0.1, which corresponds to $V_B$ between 36.39 m/s and 162.76 m/s. We conducted five integrations with the initial orbital elements of the parent body at  $a_{PB}$ = 2.3 AU, $e_{PB}$ = 0.1, $\sin(i_{PB})$ = 0.15, $\omega_{PB} = 90 ^\circ$, $f_{PB} = 45 ^\circ$, and $\Omega_{PB} = 0 ^\circ$. Consistent with \cite{delloro-2002}, we observed that $V_B$ does not alter the slope of the collision probability, but only its absolute value by a constant factor.
The impact velocities show only a weak dependence on the ejection velocity field, as illustrated in Fig. \ref{plot: impact velocity EVF dependency}. As $f_{KE}$ increases, the peak of the distribution shifts slightly toward higher impact velocities, which, however, remain well below a few hundred meters per second.

We explored the dependence on the orientation of the ejection velocity field. We assume the orbital elements of the parent body at the moment of breakup to be $a_{PB} = 2.3$ AU, $e_{PB} = 0.1$, $\sin(i_{PB}) = 0$, $f_{PB} =45^\circ$, $\omega_{PB} =0^\circ$, and $\Omega_{PB} =0^\circ$, while $f_{KE} = 0.01$. We tested for seven orientations of the symmetry axis of the ejection velocity field ($\mathbf{\vec{OC}}$ in Fig. \ref{plot: ejection velocity field}): positive tangential (parallel to the orbit), negative tangential (parallel to the orbit, but in opposite direction), radial inward (pointing to the Sun), radial outward, normal upward (perpendicular to the plane of the orbit in the N direction), normal downward, and an intermediate case with an orientation of (45$^\circ$, 45$^\circ$, 45$^\circ$) with respect to the tangential, radial and vertical directions. The results are reported in panel (d). 

We find that the orientation of the ejection velocity field does not affect the temporal decay rate of the collision probability, but it does change its absolute value by a constant factor. When the vector $\mathbf{\vec{OC}}$ is tangential to the parent body's orbit, fragments near the impact point are ejected with large velocities projected along the tangential direction, while fragments near the antipode are ejected with relatively low tangential velocities. In this configuration, the overall tangential velocity component $V_T$ is minimized compared to other EVF orientations. As follows from the Gauss equations (e.g., \citealt{zappala-1996}), for small orbital eccentricities the resulting spread in semi-major axis is proportional to $V_T$. Since a smaller dispersion in semi-major axis leads to a higher collision probability, a tangential orientation of the EVF maximizes the number of collisions. On the other hand, orientations deviating from the orbital tangent increase $V_T$, producing a more dispersed family and a lower collision probability.

Finally, we verified that the specific choice of the ejection velocity field has only a minor impact on the overall shape of the intrinsic collision probability. We performed additional tests using isotropic EVFs and the conical EVF introduced by \citet{marzari-1996} to model the Vesta family. In all cases, we found that the time evolution of the intrinsic collision probability remains unchanged.

This analysis confirms that the ejection velocity field and the parent body’s orbital elements only marginally influence the collision probability. The largest variation, about one order of magnitude, was observed when varying $f_{KE}$, which controls the magnitude of the ejection velocities. Thus, as expected, the EVF is the primary parameter affecting the intrinsic collision probability and has the strongest impact on the outcomes of the numerical simulations. In the following sections, we apply the collision probability to model the evolution in the case of an asteroid family.

\section{Generalization to real asteroid families} \label{generalization to real asteroid families}

The intrinsic collision probability that we obtained can be exploited to extend the results of the numerical simulations, limited by particle number and size, to any arbitrary size distribution with a large number of particles. In this work, we describe the size distribution as a broken power law \newtext{\citep{parker-2008}}:

\begin{align}
    f(r) &=
\begin{cases}
k \, r^{-\beta_1}, & \text{if } r > r_{{cut}} \\
k \, r_{{cut}}^{\beta_2 - \beta_1} \, r^{-\beta_2} & \text{if } r < r_{{cut}}
\end{cases}.
    \label{eq: sizedistribution}
\end{align}

This is a very general choice, permitting an adaptation to different possible outcomes of the family-forming events. In fact, while present-day asteroid family size distributions are fairly well constrained for large diameters, they are the result of a collisional erosion of the original distributions (\citealt{marzari-1995}), whose reconstruction remains difficult. 

\newtext{$r_{{cut}}$ denotes the transition radius between the gravity-dominated regime at large sizes and the strength-dominated regime at small sizes, which is typically expected to occur around 100 m \citep{donnison-2003}.}In this work, we adopt a transition radius $r_{cut} = 100$ m. This value is significantly smaller than the transition radius inferred for present-day asteroid family size distributions, which is typically on the order of $\sim$10 km \citep{morbidelli-2003, parker-2008}. The larger transition radius observed today is understood as the outcome of long-term collisional evolution with the background population. Our choice of a smaller $r_{cut}$ is therefore intended to better represent the primordial size distribution shortly after family formation.

The size distribution of Eq. \ref{eq: sizedistribution} is normalized to a fixed total mass $M$. The shape of the distribution is determined by the exponents $\beta_1$ and $\beta_2$, while $k$ is a normalization constant set to have exactly one fragment at $R_{LR}$, where $R_{LR}$ is the radius of the largest remnant ($R_{LR} \leq R_{PB}$). Primordial family size distributions can have $\beta_1$ values ranging from 2 to 8 \citep{tanga-1999, parker-2008}, whereas little information is available on $\beta_2$. \newtext{This is because most observational constraints refer to present-day size distributions, which have been modified by subsequent collisional evolution and therefore provide only indirect information on the primordial slopes. As a result, $\beta_1$ can be loosely constrained only for very young families, while essentially no constraints are available for $\beta_2$. For this reason, we explore a range of values for both parameters to cover the plausible space of initial size distributions.}

Once the size distribution $f(r)$ is defined, the expected number of collisions between particles with radii in the intervals $dr_1$ and $dr_2$ can be obtained by inverting Eq. \ref{eq:probability}:

\begin{align}
dN_{coll} (<t) = P(t)\, t\, f(r_1)\, f(r_2)\,(r_1^2+r_2^2)\,dr_1\,dr_2,
    \label{eq: collisions_diameter_range}
\end{align}

\noindent where $P(t)$ is the intrinsic collision probability. Integrating over the entire size range yields the total number of collisions expected within the family \newtext{\citep{petit&farinella-1993, o'brien&greenberg-2005}}.

If instead we focus on the number of impacts occurring on a specific reference fragment of radius $R$, the expression simplifies to

\begin{align}
    N_{coll} = P(t)\, t\, \int_{r_{min}}^{r_{max}} f(r)\, (R^2+r^2)\,dr,
    \label{eq: collisions_reference_particle}
\end{align}

\noindent where $r_{min}$ and $r_{max}$ represent the minimum and maximum radii in the size distribution. The upper limit, $r_{max}$, corresponds to the radius of the largest remnant, while the lower limit, $r_{min}$, is typically set to one meter in the following.

\begin{figure}
   \centering
   \includegraphics[width=\hsize]{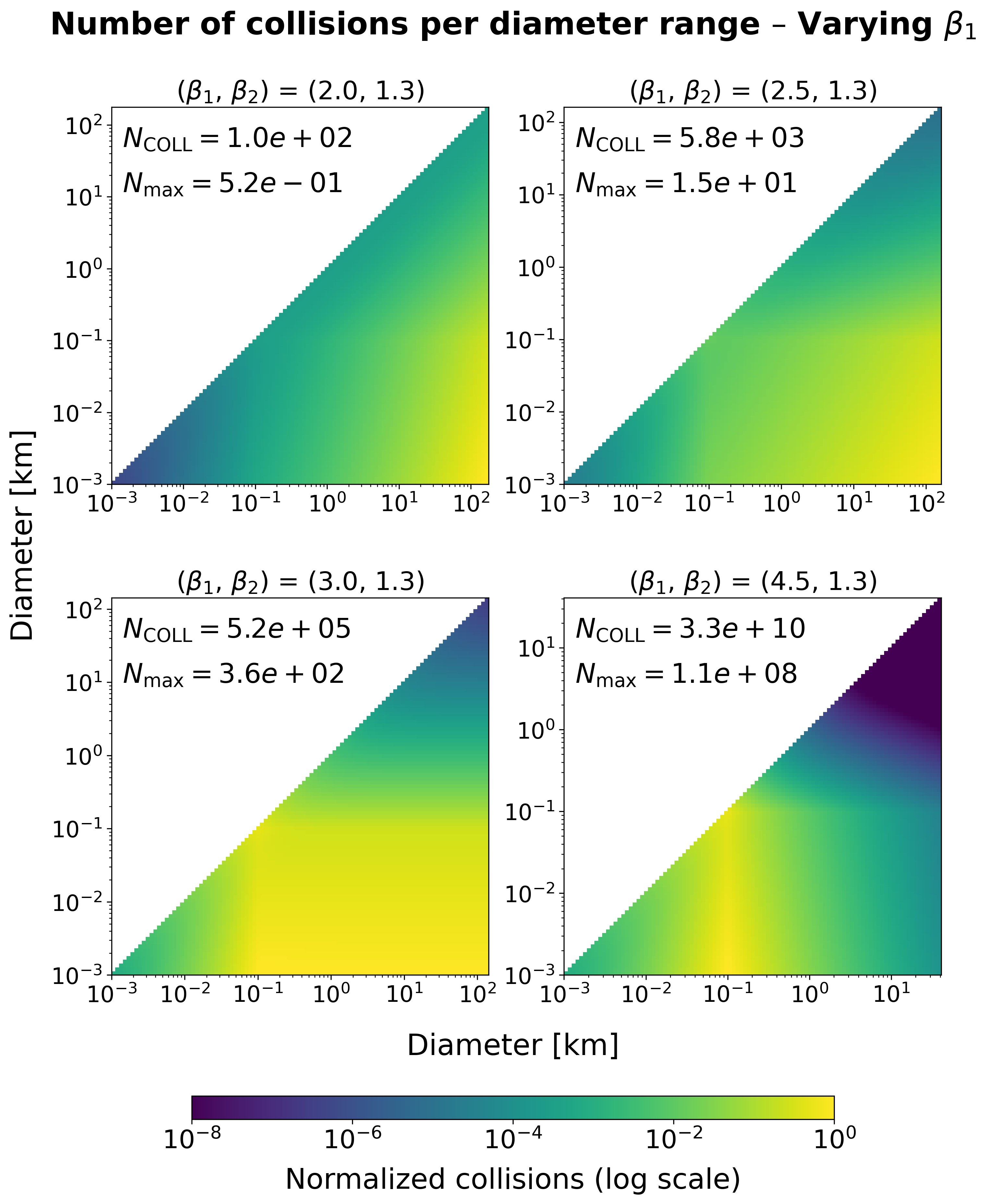}
      \caption{ Collisions per diameter range for four size distributions, varying $\beta_1$. Each panel reports the corresponding size distribution exponents and the total number of collisions $N_{COLL}$ in that distribution. Colors indicate the logarithm of the fraction of the panel maximum, with absolute maxima $N_{MAX}$ annotated in each panel. The absolute number of collisions can be recovered by multiplying the panel maximum by the value corresponding to a given color. Diameter bins were generated using 100 logarithmically spaced steps between $r_{min}$ and $R_{LR}$ for each distribution.}
    \label{plot: collision diameter range beta1}
\end{figure}

The total number of collisions within a size distribution depends on the parameters that define the distribution. The first and most straightforward dependence is on the total mass 
$M$: a larger mass implies more and larger fragments, naturally increasing the collision rate.
The same logic applies to the exponents $\beta_1$ and $\beta_2$. An increasing $\beta_1$ produces a size distribution with more small fragments, which in turn leads to more collisions. As shown by Fig. \ref{plot: collision diameter range beta1}, even a modest variation in $\beta_1$ results in a large increase in the number of collisions, from roughly $10^2$ to $>10^{10}$ for $\beta_1$=2 and $\beta_1=4.5$ respectively. A similar trend is observed for $\beta_2$, as larger values led to more particles in the distribution and thus to more collisions.

When $M$, $\beta_1$, and $\beta_2$ are fixed, the total number of collisions is inversely proportional to the radius of the largest remnant. A smaller largest remnant requires a greater number of fragments to account for the total mass, thereby increasing the overall collision rate. In summary, the total number of collisions is primarily controlled by the total number of fragments in the size distribution.

Assessing the potential role of intrafamily collisions in shaping family members’ surfaces requires determining how the number of collisions depends on the diameters of the fragments within the size distribution. To examine this correlation, we divided the distribution into small diameter intervals, $dr_1$ and $dr_2$, and calculated the total number of collisions for each combination of fragment sizes from Eq. \ref{eq: collisions_diameter_range}. \newtext{We computed the collision rates over 1,000 years. Because the intrinsic collision probability decreases exponentially, most collisions occur during the earliest dynamical phase, while the subsequent phase after the randomization of the true anomalies contributes only marginally to the total number of collisions.} \newtext{Throughout this section, we adopt the same orbital elements for the parent body as in Sect. \ref{example of a numerical simulation}.} The results are shown in Fig. \ref{plot: collision diameter range beta1}. Each panel reports the combination of ($\beta_1$, $\beta_2$) used to generate the distribution and the total number of collisions $N_{COLL}$ within it. We note that the color scale in Fig. \ref{plot: collision diameter range beta1} is normalized independently for each panel and so, the absolute collision numbers are not directly comparable across different ($\beta_1$, $\beta_2$) combinations. Larger values of $\beta_1$ lead to a substantially higher number of fragments, particularly at small sizes, thus enhancing the overall collision rate. The figure illustrates a clear trend. For small $\beta_1$s, the most likely collisions involve small particles  (on the order of one meter) impacting the largest remnant, while collisions among small fragments are rare. For intermediate exponents, such as $\beta_1 = 3$, collisions between particles with $r_{cut} < d< R_{LR}$ dominate, whereas those involving only large fragments are unlikely. Finally, for the larger exponents, such as $\beta_1=4.5$, collisions involving fragments near the transition radius $r_{cut}$ dominate, while collisions between large fragments remain the least probable. The progressive shift of the collision peak to smaller diameters reflects the changing relative abundances of fragment sizes as $\beta_1$ increases. For large $\beta_1$, the number of large fragments, above 10 km, decreases sharply, reducing the frequency of small-large collisions, while collisions involving fragments near the transition radius $r_{cut}$ become the primary contributors to the collision statistics.

\begin{figure}
   \centering
   \includegraphics[width=\hsize]{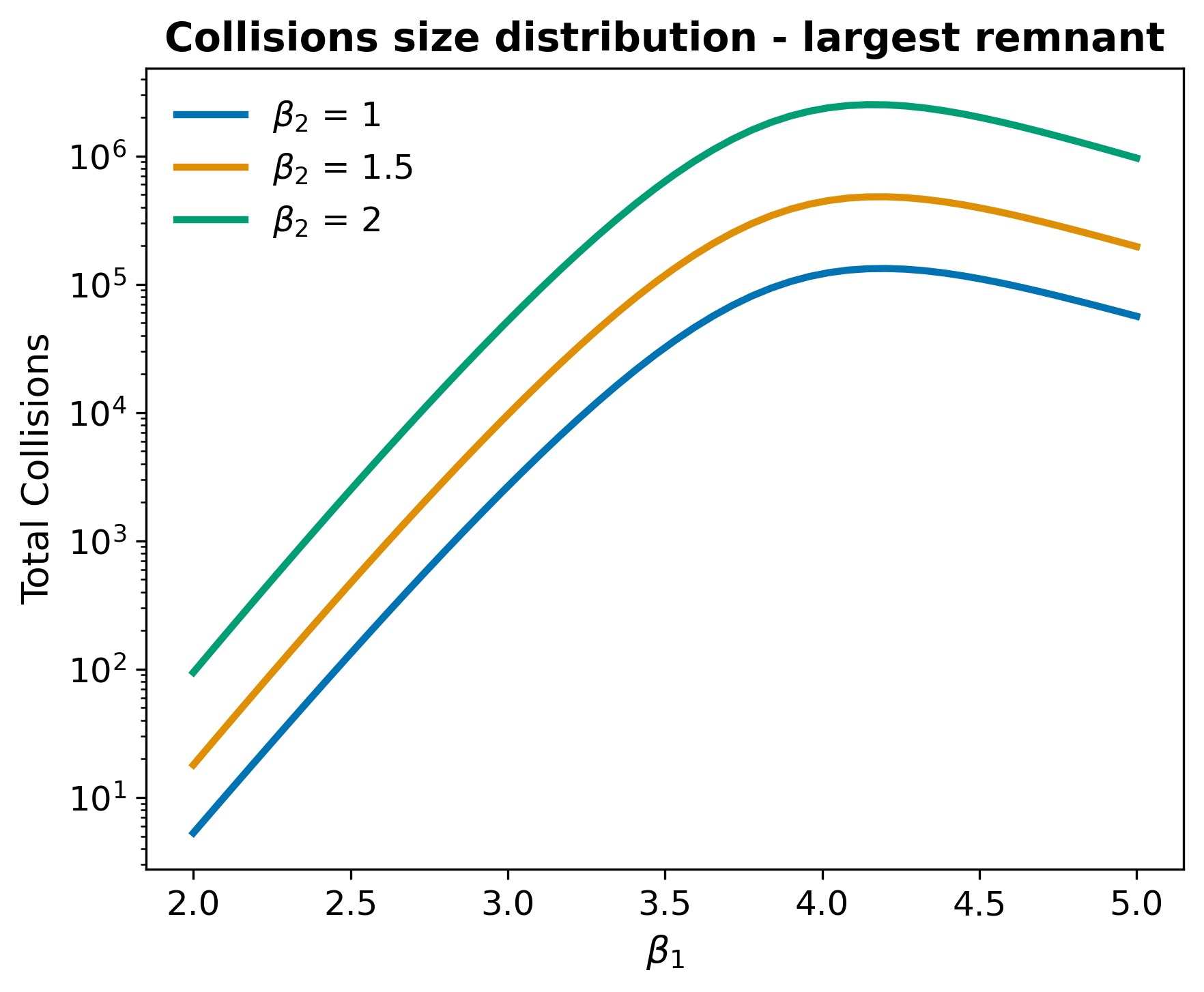}
      \caption{ Total number of collisions between fragments from different size distributions and their largest remnants. The size distributions were generated with a fixed total mass $M = 2.6 \times 10^{20}$ Kg while varying $\beta_1$ and $\beta_2$.}
    \label{plot: collision lr beta1}
\end{figure}

Typically, surface properties of family members are known only for the largest objects, thanks to spectroscopic observations. Therefore, it is particularly relevant to compute the total number of collisions on specific bodies, such as the largest remnant (LR). This can be done using Eq. \ref{eq: collisions_reference_particle}, setting $R=R_{LR}$. Figure \ref{plot: collision lr beta1}, calculated for a fixed total mass $M = 2.6 \times 10^{20}$ Kg \newtext{and over a time of 1,000 years}, shows that the total number of collisions on the LR depends on the exponents of the size distribution. While variations of $\beta_2$ preserve the slope, but change the total collision number, the dependence on $\beta_1$ is more complex. For small $\beta_1$ values, such as $\beta_1=2$, there are very few collisions on the LR. In addition, as these collisions involve small fragments on the order of one meter (see Fig. \ref{plot: collision diameter range beta1}), they are unlikely to produce large-scale modifications of the LR surface. We verified that changing the minimum fragment size $r_{min}$ in the distribution affects only the total number of collisions on the LR, without altering the overall shape of the curve.

The number of collisions increases with $\beta_1$, peaking around $\beta_1 = 4$, for which more than one million impacts can occur with fragments ranging from one to one hundred meters in diameter. Under these conditions, intrafamily collisions may significantly contribute to early surface modifications. The physical effects of these collisions are discussed further in Sect.~\ref{Importance}.

For $\beta_1 > 4$, the number of collisions begins to decrease. This occurs because the diameter of the LR decreases as $\beta_1$ increases. The reduction in its cross-section becomes more effective than the increase in the number of fragments, thereby inverting the trend.

To summarize the main results of this exploration, we find that low-velocity collisions are dominated by a swarm of small particles, so that the shape of the size distribution and the choice of its lower cut are relevant. However, Fig.~\ref{plot: collision diameter range beta1} shows that contributing fragments can extend over one or two orders of magnitude in size. In addition, the impacts on the largest family members (such as the LR) dominate for small slopes of the size distribution, but they start to diminish for the steepest distributions, where the peak of more frequent collisions starts to move towards smaller diameters. 

To assess whether intrafamily collisions may have played a significant role in the evolution of asteroid families, it is necessary to compare them with collisions between family members and the background asteroid population. If collisions with the background dominate over intrafamily collisions, the latter are likely to have played only a minor role in the evolutionary history of families.

\begin{figure}
   \centering
   \includegraphics[width=\hsize]{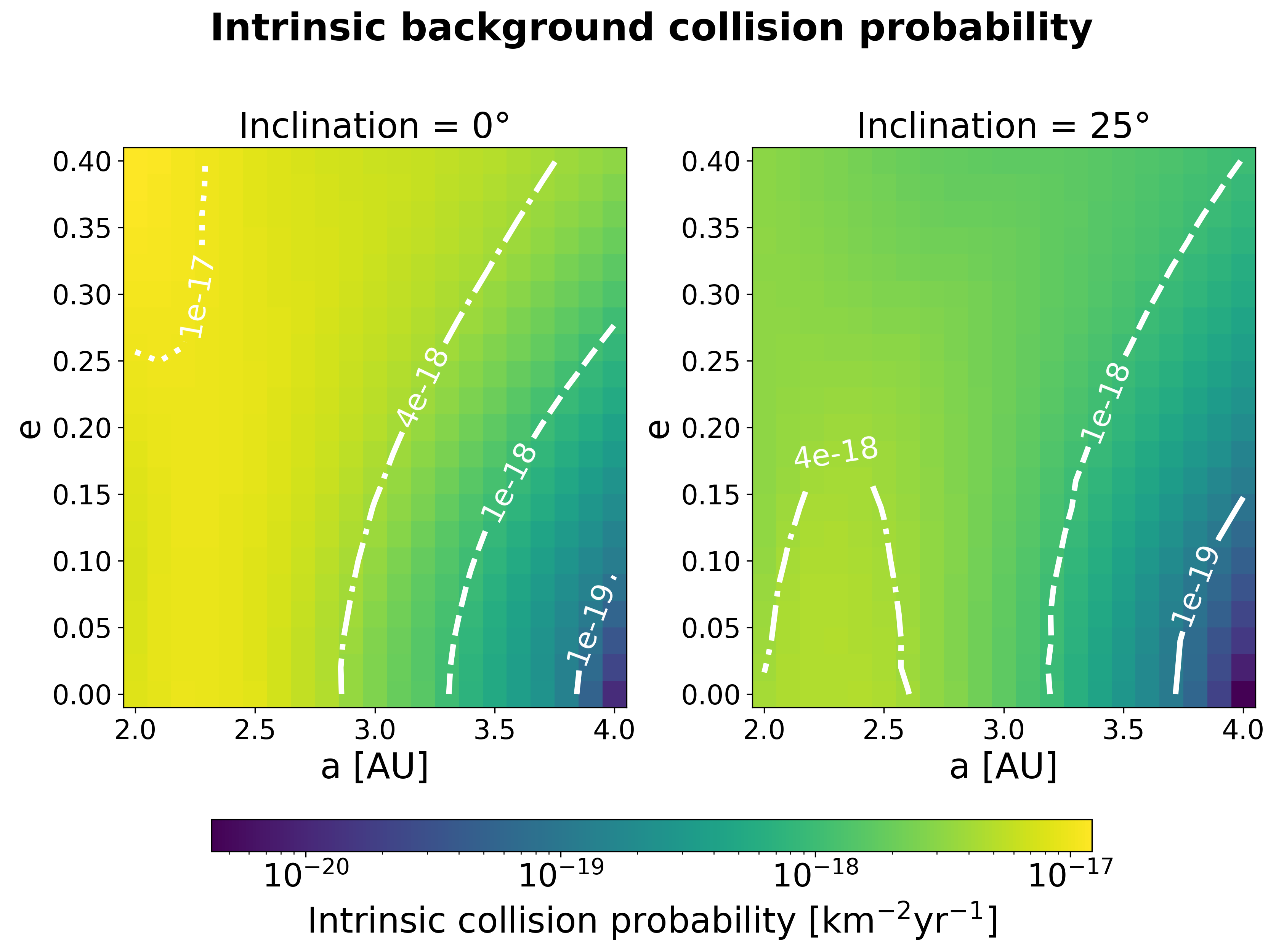}
      \caption{ Intrinsic collision probability with background asteroids across the main belt. Given a combination of ($a$, $e$, $i$) within the main
belt, the plot shows the probability of collision between an asteroid at that location and the background asteroid population. White contours indicate isolines of constant intrinsic background collision probability.}
    \label{plot: collision probability background}
\end{figure}

\begin{figure}
   \centering
   \includegraphics[width=\hsize]{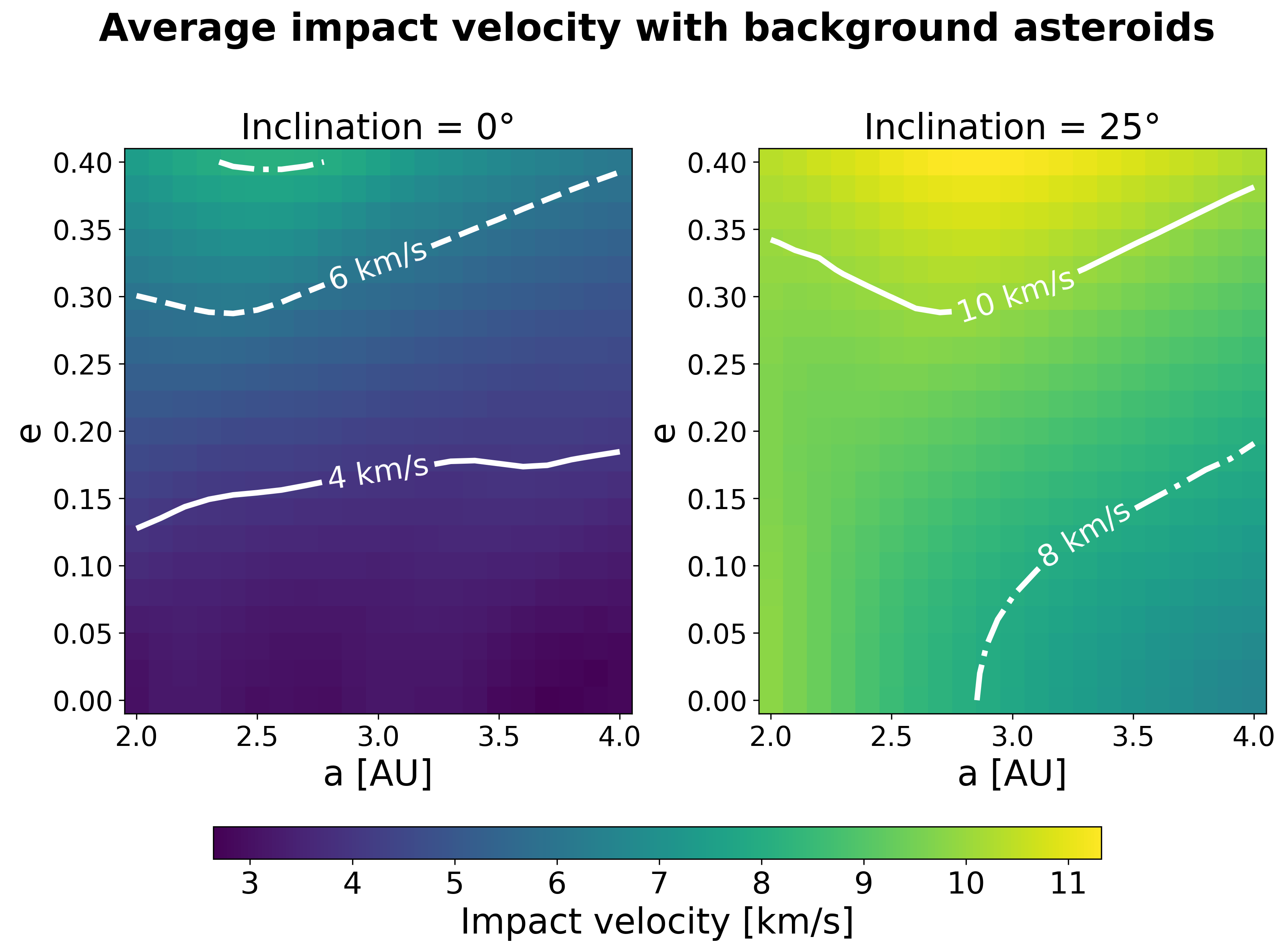}
      \caption{Average impact velocity with background asteroids across the main belt. Each panel shows the mean impact velocity between a target asteroid and the background population for different combinations of ($a$, $e$, $i$). White contours indicate isolines of constant average background impact velocity. }
    \label{plot: impact velocity background}
\end{figure}

We therefore compute the intrinsic collision probability with the background population, $P_B$. Background collisions are defined here as collisions with main
belt asteroids excluding the present-day asteroid families. The computation of $P_B$ is performed using representative projectile orbits for the background population, following the approach of \citet{delloro-2001} and \citet{delloro-2007}. We evaluate $P_B$ over a grid of semi-major axis $a$, eccentricity $e$, and inclination $i$ values representative of the main
belt. The resulting collision probabilities are shown in Fig.~\ref{plot: collision probability background}. In addition, we compute the corresponding average impact velocity with the background population, $U$, shown in Fig.~\ref{plot: impact velocity background}.

We find that the collision probability with the background population is several orders of magnitude lower than the intrafamily collision probability. However, background collisions act over very long timescales, hundreds of millions of years, comparable to typical asteroid family ages. On the other hand, intrafamily collisions are effective only over a short interval, on the order of a few hundred years following family formation. As a result, the collisional regime experienced by a family member is characterized by an initial, intense phase of low-velocity intrafamily collisions, followed by a much longer but less intense phase of collisions with the background population.

Moreover, typical impact velocities with the background population are on the order of 10 km/s, approximately three orders of magnitude higher than those of intrafamily collisions. These two collisional regimes therefore correspond to fundamentally different physical processes, with distinct consequences. Intrafamily collisions are not expected to produce further fragmentation within the family, but may instead lead to cratering or accretional merging, as discussed in the following section. Consequently, intrafamily collisions should not be disregarded a priori. Their relative importance with respect to background collisions must be assessed on a case-by-case basis, as they may be negligible for some families and significant for others.

\section{Potential physical consequences of intrafamily collisions} \label{Importance}

Intrafamily collisions occur at velocities that are too low to produce significant erosion and alter the size distribution of the family. Nevertheless, laboratory experiments (\citealt{uehara-2003}, \citealt{takita&sumita-2013}, \citealt{hayashi-2017}), and numerical simulations (\citealt{celik-2022}, \citealt{langner-2025}) show that low-velocity impacts can still excavate shallow craters. We can thus deduce that a short but intense early phase of intrafamily collisions may produce a fast early alteration on the surfaces of family members. In fact, these impacts could create a first craterization. \newtext{If the surface of a family member has already reached crater saturation, the signatures of early intrafamily collisions are likely to be erased by subsequent high-velocity impacts. In contrast, younger surfaces may still retain traces of this early phase, potentially appearing as an excess of craters and leading to a discrepancy between the geological age inferred from crater counting and the dynamical age of the family. Confirming this scenario would require high-resolution surface observations of family members from recent disruptions (such as the Datura family, 0.5-1~ Myr) and a better understanding of crater morphology at very low impact velocities.} 

Intrafamily collisions might also contribute to the early development of a regolith layer, and, if the colliding fragments differ in composition, potentially alter the surface homogeneity. In addition, very low-velocity encounters between similarly sized fragments may also result in gentle merging, producing bilobate shapes \citep{jutzi2019shapes}. While this mechanism has primarily been studied in the immediate post-breakup re-accumulation phase (\citealt{sugiura-2018}), our results suggest that such events could continue on several successive returns of the intrafamily encounter phase, for a few hundred years.

\begin{figure*}[ht]
   \centering
   \includegraphics[width=\textwidth]{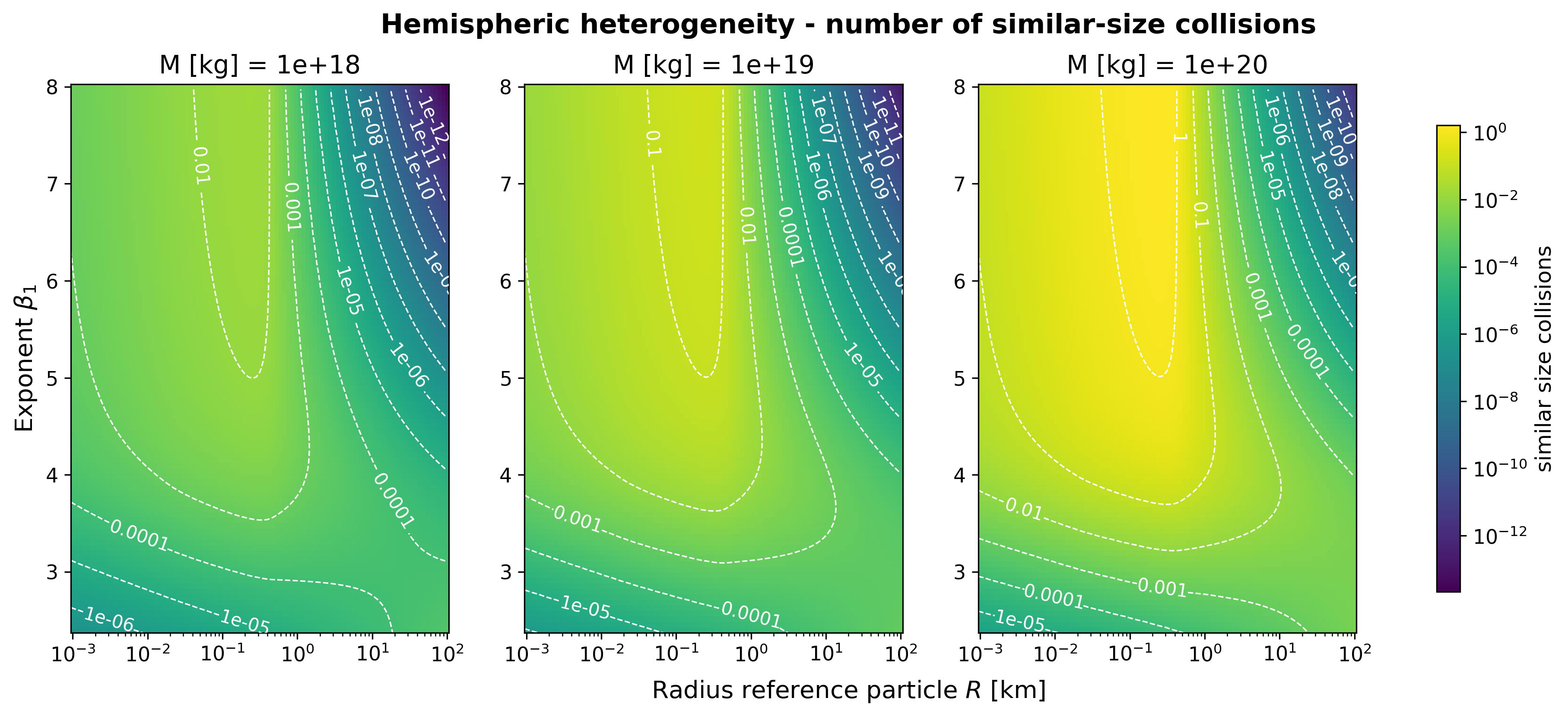}
      \caption{ Number of collisions between similar-sized fragments. Each panel shows, for different combinations of total mass $M$ and exponent $\beta_1$, the number of collisions between a reference particle of radius $R$ and the fragments with radii between $R/5$ and $5R$. White dashed contours indicate isolines of constant collision number. These plots have been produced assuming a density of 1.5 g/cm$^3$ for the parent body. }
    \label{plot: same_size_collisions}
\end{figure*}

A precise quantification of each of these processes would require specific studies and detailed numerical simulations that are beyond the scope of this work. Here we limit ourselves to the tentative comparison of our results to the observational evidence of surface heterogeneity from the spectroscopic surveys presented in \cite{hasegawa+2024}. In that study, the authors investigate hemispheric-scale surface heterogeneity in a sample of optimally observed 130 main belt asteroids by multi-epoch spectroscopy. They identified 12 very plausible candidates for variations, selected with strict and conservative criteria. They also selected 52 so-called optimistic candidates that they deemed as likely to exhibit spectral variability, with the smallest being 5 km in diameter. Considering only conservative detections, roughly 10\% of their sample displays hemispheric heterogeneity. Given the modest sample size of 130 objects, this represents a limited statistical basis, and further observations are required to quantify the heterogeneity degree across the main belt. 

In our model, hemispherical heterogeneity might be produced by low-velocity encounters between similarly sized fragments, under the hypothesis that the disrupted parent body was not totally homogeneous in composition. Such impacts can lead to partial accretion or even merger, resulting in the formation of a bilobate object. If the two lobes have different compositions, the resulting body would exhibit spectral variability. Similar outcomes may also be obtained during the immediate reaccumulation phase after the family-forming event (\citealt{sugiura-2018}). It is interesting to note that they cannot instead be produced by collisions with generic background asteroids, whose typical relative velocities of several km/s inject too much energy in the collisional process to result in gentle merging, inducing more easily massive deformation/disruption and mass loss from the involved bodies.

To evaluate whether intrafamily collisions can generate the fraction of heterogeneous objects reported by \cite{hasegawa+2024}, we used the intrinsic collision probability reported in Fig. \ref{plot: collision probability} to compute how often similar-size collisions occur within a family with a size distribution described by Eq. \ref{eq: sizedistribution}. Here, "similar-size collisions" refer to impacts between a reference fragment of radius $R$ and fragments with radii between $R/5$ and $5R$. 

The results, reported in Fig.~\ref{plot: same_size_collisions}, explore different combinations of the total mass $M$, the exponent $\beta_1$, and the radius $R$. Collisions were computed over a time $T$ = 1,000 yr, with  $\beta_2 = 2.4$, $r_{cut} = 0.1$ km, and $\rho$ = 1.5 g/cm$^3$. White dashed lines indicate isolines of constant collision number. The line labeled by 0.1 shows the combination of ($R$, $\beta_1$) for which one in every ten fragments undergoes a similar-size impact, matching our reference rate. Our results depend weakly on the assumed density, since for fixed $M$ and $\beta_1$, a higher density implies fewer fragments in the family and thus a slightly lower collision rate.

Our results show that similar-size collisions can create hemispheric heterogeneity for $M \geq 10^{19}$ kg, but primarily for sub-kilometer objects, for which high-resolution multi-epoch spectroscopy is not currently available. For initial masses $M \geq 10^{20}$ kg, intrafamily similar-size collisions could heterogenize nearly all sub-kilometer fragments. In contrast, for larger asteroids like the ones observed by \citet[$R$ > 10 km]{hasegawa+2024}, such collisions become exceedingly rare: even at high initial masses, only about one in a thousand bodies would experience a similar-size encounter capable of producing hemispheric heterogeneity. \newtext{Variations in other size-distribution parameters or breakup conditions may slightly increase the collision probability, but cannot reproduce the observed fraction of heterogeneous objects.}

\newtext{Large-scale heterogeneity in larger family members ($R > 10$ km) may instead result from repeated impacts with smaller fragments. Large bodies can experience millions of impacts from meter- to hundred-meter-sized projectiles (Fig. \ref{plot: collision lr beta1}). While individual impacts would not produce detectable spectral signatures, their cumulative effect could cover significant surface areas.}

\newtext{Whether this bombardment can generate observable heterogeneity depends on whether impacts preferentially occur in specific regions. Anisotropic impact patterns are possible even when impactors arrive from random directions, for a rotating ellipsoidal target \citep{liu-2025}. The orbital geometry of family members may further concentrate impacts near specific surface areas. If impactors differ in composition from the target, such anisotropic bombardment could potentially generate observable surface heterogeneity. However, testing this scenario would require dedicated numerical simulations, which are outside the scope of this study.}

\newtext{The comparison with hemispheric spectral heterogeneity here presented relies on several assumptions about the size distribution, the total family mass, and the adopted timescale, and should therefore be regarded as illustrative rather than predictive.}

\section{Conclusions} \label{Conclusion}

In this work, we investigated intrafamily collisions occurring in the early dynamical phases following the formation of an asteroid family. We performed numerical simulations tracking the evolution of fragments up to the randomization age of the true anomalies, recording mutual impacts and converting them into an intrinsic collision probability. We show that the collision probability can reach values as high as  $10^{-10}$ yr$^{-1}$km$^{-2}$ in the first years after breakup, after which it decreases exponentially, following the same trend predicted by the statistical method of \cite{delloro-2002}. In practical terms, we show that the first $\sim$1,000 years of the evolution of a main belt family produce two orders of magnitude more intrafamily collisions than the $\sim$10,000 years following the randomization of the true anomalies. 

We examined how the collision probability depends on the simulation parameters. We observed that variations in the orbital elements of the parent body and in the properties of the ejection velocity field can modify it up to one or two orders of magnitude, but its temporal evolution is not affected.

We then developed a statistical method to extend the results of our numerical simulations to arbitrary family size distributions, modeled as broken power laws. We explored how the number of collisions depends on the parameters defining the size distribution. As expected, for shallow size distributions, collisions are dominated by small projectiles impacting the largest fragments, whereas for steeper distributions, collisions among sub-kilometer fragments become the most frequent.

We find that intrafamily collisions occur at very low relative velocities, typically between a few meters per second and at most a few hundred meters per second. As already noted by \cite{delloro-2002}, such low-speed impacts are not energetic enough to significantly alter the overall size distribution of the family. 

Most observations of physical properties are typically available only for the larger members of families, so our analysis focused mainly on the largest remnant. Depending on the assumed size distributions, the number of impacts ranges from less than ten to several million. In the latter cases, intrafamily collisions could produce a primordial cratering record and regolith layer on the surface of the largest remnant. The role of such early surface gardening should be further studied to assess its relevance in different scenarios.

We also investigated how low-velocity intrafamily collisions could generate large-scale heterogeneities in family members. We showed that sub–kilometer fragments are frequently involved in collisions with similarly sized fragments. At such low relative speeds, the most probable outcome is accretion rather than catastrophic disruption, producing bilobate remnants. Moreover, if the two fragments have different surface compositions, the merged object could show hemispherical heterogeneity. A direct observational verification remains difficult due to the lack of rotationally resolved spectroscopy for small family members. \cite{hasegawa+2024} report that roughly 10\% of large main
belt asteroids exhibit hemispherical heterogeneity. Although equal-size intrafamily collisions are unlikely for large objects, our model predicts that the largest family members may experience a large number of impacts from small fragments.

We highlight here some limitations of this work. First, the lack of strong constraints on the post-breakup size distribution of asteroid families makes it difficult to finely tune the simulations. Also, we adopted the semi-empirical ejection-velocity
field of \cite{paolicchi-1996}, as a realistic starting point to model the structure of the ejection initial conditions. SPH models produce a more complex ejection velocity field
than the SEM \citep{delloro-2015}. However, the collision probability is primarily controlled by the resulting spread in semi-major axis, which depends on the magnitude of the overall dispersion of the ejection velocities rather than on the detailed structure of the EVF itself. Consequently, no major differences are expected if SPH-based EVFs are adopted instead of the SEM, although this assumption should be tested with more detailed simulations. 

Finally, we do not take into account inter-particle gravity among family members. Indeed, mutual gravitational scattering among family members can occur only near the common nodes of their orbits and could slightly enhance the dispersion of small fragments interacting with the largest ones. The role of this effect, probably of secondary relevance, could be studied in more detail in future works.

In conclusion, our results extend the study of \citet{delloro-2002} to the earliest phases of asteroid family formation. We find evidence for the possible existence of an early intrafamily phase of low-velocity collisions, whose intensity depends on the details of the ejection velocity field and the size distribution. We do not claim that intrafamily collisions can account for the collisional evolution and surface processing of all asteroid family members; rather, they represent an additional mechanism that may play a significant role in some families while being negligible in others. Future investigations will help clarify the relevance of this process in different dynamical and physical scenarios.

\begin{acknowledgements}

We acknowledge useful discussions with several colleagues, in particular A. Cellino, A.C. Bagatin, P. Paolicchi. 

RB Doctoral contract is funded by Universit\`e de la C\^ote d'Azur.

ADO has been supported by the call 2023 of the Italian National Institute for Astrophysics for fundamental research projects (INAF, act n. 38/2023).

This project was financed in part by the French Programme National de Planetologie, and by the BQR program of Observatoire de la C\^ote d'Azur. It was also supported by the French government through the France 2030 investment plan managed by the National Research Agency (ANR), as part of the Initiative of Excellence Université Côte d’Azur under reference number ANR-15-IDEX-01. The authors are grateful to the Université Côte d’Azur’s Center for High-Performance Computing (OPAL infrastructure) for providing resources and support.

We made use of Astropy, a community-developed core Python package for Astronomy \citep{0astropy2013, 1astropy2018, 2astropy2022}; Matplotlib \citep{matplotlib_Hunter:2007}.

\end{acknowledgements}

\bibliographystyle{aa}
\bibliography{references}

\end{document}